\documentclass[utf8]{FrontiersinHarvard} 

\usepackage{url,hyperref,lineno,microtype}
\usepackage[onehalfspacing]{setspace}
\hypersetup{
  hidelinks
}

\def\keyFont{\fontsize{8}{11}\helveticabold }
\def\firstAuthorLast{Roman and Bertolotti} 
\def\Authors{Sabin Roman,$^{1,2,3*}$, Francesco Bertolotti,$^{3,4}$}
\def\Address{$^{1}$ Department of Knowledge Technologies, Jo\u{z}ef Stefan Institute, Ljubljana, Slovenia\\
$^{2}$ Center for the Study of Existential Risk, University of Cambridge, Cambridge, UK\\
$^{3}$ Intelligence, Complexity and Technology Lab (ICT Lab), LIUC – Carlo Cattaneo University, Castellanza, Italy\\
$^{4}$ School of Industrial Engineering, LIUC – Carlo Cattaneo University, Castellanza, Italy
}

\begin{document}
\onecolumn
\firstpage{1}

\title[Toward a new AI winter?]{Toward a new AI winter? How diffusion of technological innovation on networks leads to chaotic boom-bust cycles} 

\author[\firstAuthorLast ]{\Authors} 
\address{} 
\correspondance{} 
\extraAuth{}

\maketitle

\begin{abstract}

Technological developments and the impact of artificial intelligence (AI) are omnipresent themes and concerns of the present day. Much has been written on these topics but applications of quantitative models to understand the techno-social landscape have been much more limited. We propose a mathematical model that can help understand in a unified manner the patterns underlying technological development and also identify the different regimes in which the technological landscape evolves. First, we develop a model of innovation diffusion between different technologies, the growth of each reinforcing the development of the others. The model has a variable that quantifies the level of development (or innovation, discovery) potential for a given technology. The potential, or market capacity, increases via diffusion from related technologies, reflecting the fact that a technology does not develop in isolation. Hence, the growth of each technology is influenced by how developed its neighboring (related) technologies are. This allows us to reproduce long-term trends seen in computing technology and large language models (LLMs). We then present a three-dimensional system of supply, demand, and investment which shows oscillations (business cycles) emerging if investment is too high into a given technology, product, or market. We finally combine the two models through a common variable and show that if investment or diffusion is too high in the network context, chaotic boom-bust cycles can emerge. These quantitative considerations allow us to reproduce the boom-bust patterns seen in non-fungible token (NFT) transaction data and also have deep implications for the development of AI which we highlight, such as the arrival of a new AI winter.

\tiny
 \keyFont{ \section{Keywords:} AI winter, artificial intelligence, innovation diffusion, technology, chaos, networks, simulation, market bubble} 
\end{abstract}

\section{Introduction}

In recent years, studies on innovation diffusion and technological development have increasingly focused their attention to artificial intelligence (AI), treating it as both a novel technological trajectory and a test case for established theoretical frameworks \citep{gherhes2023technological, bertolotti2025llm}. While foundational models, such as Rogers’ diffusion curve or theories of technological paradigms, offer useful starting points \citep{rogers1962diffusion}, AI presents distinct challenges \citep{dahlke2024epidemic}. Its general-purpose character, rapid and recursive advancement, and interdependency between data infrastructure, computational power and communication protocols complicate traditional assumptions about how technologies spread and evolve \citep{vargas2025artificial}. The diffusion of AI is highly uneven \citep{madan2023ai}: rather than following a single trajectory, AI development and uptake unfold across multiple domains simultaneously, with feedback loops between users, firms, and technical communities \citep{christian2024applying}. Pre-existent dynamical models cannot adequately capture these features: they assume relatively independent trajectories of adoption or linear market dynamics, different from AI's diffusion \citep{rogers1962diffusion, bass1969new, hung2016innovations, zhang2023evolution}. Given these novel dynamics that AI development poses, we develop new models of innovation diffusion that take these elements into consideration and are applicable both to quantifying long-established trends within the computing industry and to more recent developments.

In the theory of innovation diffusion and its models, one of the focus is on individual consumers that buy and use the products and how diffusion occurs in spreading awareness of the goods \citep{singh2022search}. In the present study, we depart from this interpretation and try to understand how technological innovation -- considered as an expansion of the dimension of a technological system other that the mere changes of its parameters, that could be perceived by users and buyers as something new and worth their interest \citep{dejong2025socio}, and consequently lead to its diffusion -- can spur other technologies to grow and develop in the same manner. For example, the way it happened with the introduction of the transistor, which catalyzed a whole host of other technologies across the computing industry and ultimately across all industries. Especially, with the development of each technology, the demand grows for related technologies, either to be created or refined. This cycle can facilitate rapid, even exponentially fast developments, such has happened with computing technology, see Fig. \ref{fig:1}(a). We consider a deeper understanding of these phenomena crucial to assess the behavior of the AI market, with the explicit purpose of understanding under which conditions a new AI winter could happen.  

We begin with a model that aims to capture the dynamics of innovation diffusion between technological products. In addition to this mechanism we also aim to capture business dynamics, such as the allocation of investment and the production of the technologies. In this regard we introduce a second model that reproduces the well documented pattern of business cycles \citep{BENHABIB1979421,sterman2010business,bertolotti2022risk}. Finally, we merge the models into a more complete representation of the market. A surprising feature of the complete model is the emergence of transient chaos, manifested through boom-bust dynamics in conditions of high diffusion or investment. 

These dynamics matter for understanding the current status and the future of AI. The model shows that uninterrupted market expansion can engender perceptions of boundless potential, thereby precipitating overinvestment, speculation, and fragility. In the event of disillusionment, regulation, or stagnation, there is a risk that these cycles may trigger a new AI winter. The implementation of policy mechanisms to moderate investment, support openness, and provide counter-cyclical funding has the potential to stabilise long-term trajectories. 

The structure of the paper is as follows. It begins with a review of the relevant literature, followed by a detailed description of the methodological approach. Each model is then introduced alongside its corresponding results, with the findings from the initial models motivating the development of the subsequent ones. The paper concludes with a comprehensive discussion of the results and the final remarks.

\section{Literature review}

The theory of innovation diffusion is often framed and focused on the adoption of a technology by consumers \citep{vishwanath2003}, wherein a product is introduced, adoption grows, matures and eventually declines. The Bass model \citep{bass1969new} closely reflects the introduction and growth phases posited by the theory by specifying the rate of adoption. The are different extensions of the model, with some focused on capturing the maturing and decline phases \citep{mahajan1996timing}, the impact of geography on the diffusion \citep{lengyel2020role, di2024bass} or the different awareness consumers might have of the good \citep{wang2006mathematical, fibich2023diffusion}.

Innovation diffusion can be driven both by endogenous and exogenous elements \citep{gasparin2019thinking}. Regarding endogenous effects, the most relevant are feedback and product misperception \citep{paich1993boom}, cultural barriers \citep{Rolfstam2011Public}, the stock of human capital \citep{romer1990}, and rational herding derived from over-optimistic perspectives \citep{schaal2023herding}, which can alone lead to aggregate fluctuations and boom and bust cycles \citep{schaal2023herding}. On the other hand, exogenous elements include external disruptive elements \citep{roman2023theories}, such as the Covid-19 pandemic \citep{Jin2021The}, natural calamities \citep{Miao2013Necessity}, or wars \citep{Brunk1981The, roman2022master}.

As we will show later, combining innovation diffusion with business cycle dynamics can lead to the emergence of chaos. Chaos in data on technological innovation, such a time series of patents, is usually determined by means of local Lyapunov exponents \citep{hung2016innovations, zhang2023evolution}. The use of patent data has revealed chaotic dynamics underlying industries such semiconductors, software and biotechnology \citep{hung2011technological}. Business cycles \citep{kondratieff1979long} have long been studied in connection to the spread of innovation \citep{jenner1991technological, knell2022tools} and how chaos can emerge from their interaction \citep{houchin2005complexity, stengers2018order}. Prior empirical research has analyzed the role of the organizational feedback loop in the emergence of chaos in an innovation market \citep{hung2016innovations}, showing that asynchronous updating of the innovation R\&D strategy can prevent the system evolution from turning into chaos \citep{zhang2023evolution}. The market's information flow is often volatile in a way that challenges conventional modeling assumptions \citep{brody2011modelling}. Conversely, a fractal statistical-analysis approach, grounded in chaos theory \citep{roman2021historical}, can effectively provide a better understanding the market's nonlinear fundamental characteristics \citep{ke2023cross, bertolotti2023methodology}.

A possible example of this is the Cournot duopoly \citep{cournot1838recherches}, where, if the two firms has a sufficiently large cost differential, the junction of the two non-linear optimal responses would lose stability and chaos could emerge in the long term \citep{puu1991chaos}, but chaotic behavior could be avoided in the specific case where both firms share the same cost functions and owns no prior expectations \citep{lamantia2022discontinuous}. The chaos in such system arises from the increased sensitivity to behavioral parameters (i.e., the production function) of the two firms \citep{bischi1999symmetry}.

The spread of technological innovation is also affected by the topological structure of the market network \citep{choi2010role}, such as it happens with other diffusion phenomena \citep{dunbar2020structure}. This effect seems to be moderated by the strength of social interactions \citep{delre2010will, Kocsis2011Competition}, the existence of clusters \citep{Kocsis2011Competition, Bohlmann2010The}, and network properties such as centrality \citep{Sisodiya2012Enhancing, Chuluun2017Firm} and density \citep{Chuluun2017Firm}. The phenomenon has also been studies also by means of computer simulation, especially to gain a better understanding of feedback loop effects \citep{abrahamson1997social} and to take into consideration the effect of individuals heterogeneity in the innovation diffusion \citep{Bohlmann2010The}.

Nevertheless, an agreement does not exist regarding the specific effect of topology. This could depend on two reasons. First, the same phenomenon was investigated by very different scientific communities \citep{Sznajd-Weron2013Rewiring}. As an example, recent work \citep{arieli2020speed} proposes a game-theoretical analytical analysis where a small star-shaped network is analyzed. This approach differs sharply from the one of the complex network community \citep{newman2003structure}, that usually considers the Erdos-Reny structure \citep{erdos1960evolution} as a baseline and compares it with structure more adherent to real social networks \citep{mccullen2013multiparameter, bertolotti2024balancing}, such as small-world \citep{watts1998collective} or highly-clustered \citep{newman2003properties} networks. Second, technological innovation is a broad topic that could include elements different from each other, even from a structural point of view \citep{gasparin2019thinking}. Given that, it is not surprising that different results emerge.

Despite the growth of AI applications across sectors, the literature on the dynamics of AI diffusion remains surprisingly underdeveloped. While conceptual frameworks abound, only a limited number of studies offer formalized or quantitative diffusion models specific to AI technologies \citep{Heim2025AIFramework}. The Bass model was used to explore the diffusion of AI-related patents across the U.S. innovation ecosystem, integrating topic modeling to measure the technological maturity of various AI domains over time \citep{lee2020study}, showing that AI technologies exhibit heterogeneous life cycles, with adoption trajectories highly sensitive to domain-specific incentives.
Rogers’ diffusion of innovation theory has also been adapted to the context of AI to investigate the adoption of AI across four organizational groups—administrators, faculty, staff, and students—within a U.S. higher education institution \citep{phillips2025artificial}, showing that AI adoption outpaces classical diffusion benchmarks by a factor of 2.34. A similar epidemic-based spreading model has been implemented using firm-level data, studying over 380,000 enterprises in Germany, Austria, and Switzerland to identify adoption mechanisms for AI and other digital technologies \citep{dahlke2024epidemic}. Finally, a System Dynamics model of AI adoption in a specific context (a financial company) also exists \citep{kumari2021system}.


\section*{Methods}

In this work we introduce and study the behaviour of three models aimed at understanding the diffusion of technological innovation. The study is theoretical in nature, focused on exploring the dynamics of the innovation system, however we do aim to capture some broads trends seen in empirical data, see Fig. \ref{fig:1}(a), (b) \footnote{These graphs include information on microprocessors \citep{rupp2021microprocessor} and LLMs \citep{villalobos2024willwerunoutofdata, senthilkumar2024can}, and are either generated independently or reproduced under CC BY 4.0. The NFTs data consists of 6.1 million transactions collected mainly from the Ethereum and WAX blockchains between June 23, 2017, and April 27, 2021 \citep{nadini2021mapping}.}. The main emphasis is on the dynamics of technology and technological products with the aim of determining the regimes where innovation diffusion and market mechanisms provide a stable and reliable trajectory or where the opposite might occur. 

\subsection{Model overview}

The three models introduced study the diffusion of technological innovation in different and progressively more sophisticated ways.
The first model focuses on technological diffusion with the aim to capture the basic features seen in the development of computing technology and related sectors. A core aspect of the model is assumption that the dependence between difference technologies can be captured by a network, which is a versatile tool in modeling socio-technical relationships \citep{roman2017topology, roman2023global}. 
The second model aims to quantify the business cycle that a given product undergoes \cite{hallegatte2008business}. In this model, supply, demand and investment are the variables of interest and the model is meant to be general, showcasing the economic dynamics for any given product. The specific functional forms in the model likely restrict it in practice to certain industrial sectors, however its overall architecture is aimed to be prototypical and potentially adapted for any problem of interest. 
Finally, the third model combines the previous two models to provide insights into the market dynamics of technologies or products.

For the network models for simplicity we assume that nodes are homogeneous, namely the parameter values are the same throughout the network, and we only focus on the emergent dynamics due to the network features \citep{bertolotti2024balancing, bertolotti2025gravity}. In the present study we report the results for random regular graphs, but we have checked the robustness of our findings across different degree distributions (star, scale-free) and confirm consistency. The provided code includes options for different topologies and allows for immediate testing.

For each model, we present and contrast the main dynamical regimes; we identify the key parameters and their values that characterize each behavior. Where appropriate, we compute the bifurcation diagrams when varying the critical parameters. The models we build are dynamical systems expressed as systems of ordinary differential equations. These modeling tools are appropriate to capture causal mechanisms \citep{sterman2010business} and how they manifest over time in the behavior of the variables of interest, such as the supply, demand and investment stocks. We hope that the gradual, piece-wise introduction of the different components of the model, namely the pure diffusion, then the business cycle and finally the full market structure, aids the communication and understanding of the overall dynamics.

\subsubsection*{Pure diffusion model}

Technological innovation follows a exponential growth trajectory over time until a saturation point is reached \citep{mcnamee2017modeling}. At the saturation stage the performance metrics, production or demand for technology stagnate. New technologies, products or increased demand (e.g., from marketing or population growth) can catalyze the development of older technologies and lead to a new growth period \citep{yli1998network}.

This trend can be observed in the computing technologies, see Figure \ref{fig:1}(a), where the typical power, frequency and single thread performance are seen to have reached a plateau (or significant slowdown from the previous decades). The transistor count however has avoided this plateau due to an increase in the number of cores. Even recent development such as LLMs are susceptible to this saturation dynamic, see Figure \ref{fig:1}(b).

So, we notice a dynamics where: (1) technological growth is monotonic but not strictly as it (2) can reach a plateau and (3) development in related technologies can stimulate growth again. Given the dynamics described in (1) and (2) we propose that a given technology grows logistically:
\begin{equation}
\begin{aligned}
\dot y_{i} &= ry_{i}\left(1-\frac{y_{i}}{u_{i}}\right)\\
\dot u_{i} &= \sigma \mbox{Relu} \left(\sum_{\mbox{neighbours}\;k} y_{k} - y_{i}\right)
\label{eq:1}
\end{aligned}
\end{equation}

In equation \eqref{eq:1} the variable $y_{i}$ can represent a given metric of performance, the volume of production or the demand for a given technology $i$ (or product), while the variable $u_{i}$ is the carrying capacity of the market for the given technology. The dynamics of $u$ in \eqref{eq:1} is given by a diffusion mechanism that only promotes non-negative growth rates so that technological development remains monotonic, as seen in Fig. \ref{fig:1}. The function Relu$(x) = x$ if $x \geq 0$ and zero otherwise.

The key parameters in the model are the individual growth rates $r_{i}$ and diffusion rates $\sigma_{i}$ of the different technologies. Figures \ref{fig:1}(c) and (d) show the interplay and contrasting dynamics of different rates of growth and diffusion. For simplicity, we choose a common value for $r_{i} = r$ and $\sigma_{i} = \sigma$. The dependence between technologies is represented through an undirected graph wherein the neighbors of a given technology are the closest related/impacted technologies. The results shown in Figure \ref{fig:1} are for a random regular graph which we consider to be representative of the dependency between different technologies \citep{feray2018weighted}. The network has $N = 10$ nodes, with mean degree equal to 5, which means each technology depends on average on the five closest related products. Similar results hold for other values of $N$ or of the average degree.

\subsubsection*{Business cycle model}

Beyond the technological dynamics we focused on in the previous section we need to account for economic factors that can impact innovation. So, we propose a model of economic dynamics consisting of a system of differential equations for the supply $x$, the demand $y$ and the investment $z$. Existing models in both the business model of management domain and the business cycle modeling literature are not well-suited to our goals. Models related to Industry 4.0 and 5.0 are often highly specific or overly complex \citep{lyneis2000system, degres2004simulation, manenzhe2021maintenance, medoh2022future, velandia2023driving}, and do not reproduce emergent business cycles. Conversely, models explicitly focused on business cycles tend to be structurally rich and technically intricate, whereas our aim was to develop the simplest possible model that still captures cyclical behavior \citep{chow2009analyzing, taniguchi2009business, billio2017dynamical, kroujiline2021endogenous}.

The supply $x$ and investment $z$ are directly observable variables as the monetary costs and the production of goods are tangible. However, the demand $y$ is not generally known and is often unpredictable; there are innumerable examples of product launches that have failed due to low demand \citep{victory2021common}, whereas there is no generally established recipe for best-sellers in any market \citep{bhardwaj2023ecosystem}. Efforts to replicate past successes by formulaic approaches often fail, with numerous examples from the movie industry, book sells, games, phones, PCs and other products \citep{stokely2005success}. A product launch is often a gamble and success (e.g., is in high demand) depends on consumer preferences, existing niches and products, company reputation, marketing and investment strategy \citep{hultink2000launch}.

The business cycle model is given by:
\begin{equation}
\begin{aligned}
\dot x &= bz - dx\\
\dot y &= ry_{i}\left(1-\frac{y_{i}}{u_{0}}\right) - \alpha_{1} x y\\
\dot z &= \alpha_{2} x y - cz
\label{eq:2}
\end{aligned}
\end{equation}

The supply $x$ is affected by two factors: the $bz$ term implies that supply grows in proportion to investment, while the $-dx$ term means that a supply boom leads to a subsequent decrease. While the demand $y$ is not generally known, we can assume that the considered good is perishable or it can be consumed, so the demand recovers over time due to a replenishment need. A simple example is food consumption: after a meal the demand for food is fully satisfied (so the demand drops to zero), whereas after a time the demand recovers as we get more hungry. This dynamic can also be seen in the cyclic production of goods in series (e.g., cars), or generations (e.g., game consoles) or versions (e.g., software) \citep{beraja2021demand}.

The demand $y$ satisfies an equation incorporating logistic growth and a simple interaction such as the one appearing in the Lotka–Volterra  predation model. The logistic term $ry(1-y/u)$ means that demand recovers over time (if no supply satisfies it). The term $- \alpha_{1} x y$ implies that the demand is met (so it decreases) to the extent that supply encounters demand. If a product $x$ has no visibility to consumers (e.g., poor marketing) then it unknown and cannot meet demand, hence $\alpha_{1} = 0$. Similarly, if supply $x$ is low then demand is only slowly met. There are numerous possible choices of functional forms for the interaction between supply and demand \citep{huang2013demand}, however the predator term $xy$ is one of the simplest options that gives realistic dynamics (such as oscillations), while also taking into consideration the existence of relationships between the modelled entities (buyers and sellers).

The investment $z$ equation also has two parts. The term $\alpha_{2} x y$ signifies that investment increases if demand $y$ is high or the market is producing $x$ in large volumes. Large investments are followed by a decrease \citep{demirhan2005information}, hence the term $-cz$. In general $\alpha_{1} \neq \alpha_{2}$ however for simplicity we assume equality $\alpha_{1} = \alpha_{2} = \alpha$. The rationale for this is that once a product successfully engages with consumer demand (and so becomes profitable), then there's a strong market signal to increase investments in the product in proportion to how much demand is being met \citep{condorelli2025fair}, i.e. proportional to $xy$. For example, the success of the iPhone spurred a huge investment and market increase in smart phones \citep{mallinson2015smartphone}.

\subsubsection*{Techno-landscape market model}

Finally, we combine the two previous models to obtain a dynamical system more adherent to how related technologies evolve in response to each other and how market signals affect the overall behavior of the network. The complete model is given by:
\begin{equation}
\begin{aligned}
\dot x_{i} &= bz_{i} - dx_{i}\\
\dot y_{i} &= ry_{i}\left(1-\frac{y_{i}}{u_{i}}\right) - \alpha x_{i} y_{i}\\
\dot z_{i} &= \alpha x_{i} y_{i} - cz_{i}\\
\dot u_{i} &= \sigma \mbox{Relu} \left(\sum_{\mbox{neighbours}\;k} y_{k} - y_{i}\right)
\label{eq:3}
\end{aligned}
\end{equation}

The supply $x_{i}$ and investment $z_{i}$ dynamics is the same as \eqref{eq:2} applied to each node $i$ (which is a technology or a product). The demand and market capacity satisfy the dynamics of \eqref{eq:1} except that now demand can be satisfied by supply through the term $-\alpha x_{i} y_{i}$ term.

\subsection{Experimental design}

The models were implemented in Julia with the DifferentialEquations library (version 7.16.1). The decision regarding the programming code depends on the possibility of facilitating faster computation than other popular languages such as Python, providing a variety of libraries suitable for scientific computing \citep{bezanson2017julia}, and at the same time maintaining high readability, which is important for replicability. We solve the models using the 4th-order Runge-Kutta method with a time step of $0.25$.

\begin{figure}[t]
\centering

\begin{minipage}{0.45\textwidth}
  \centering
  \includegraphics[width=\linewidth]{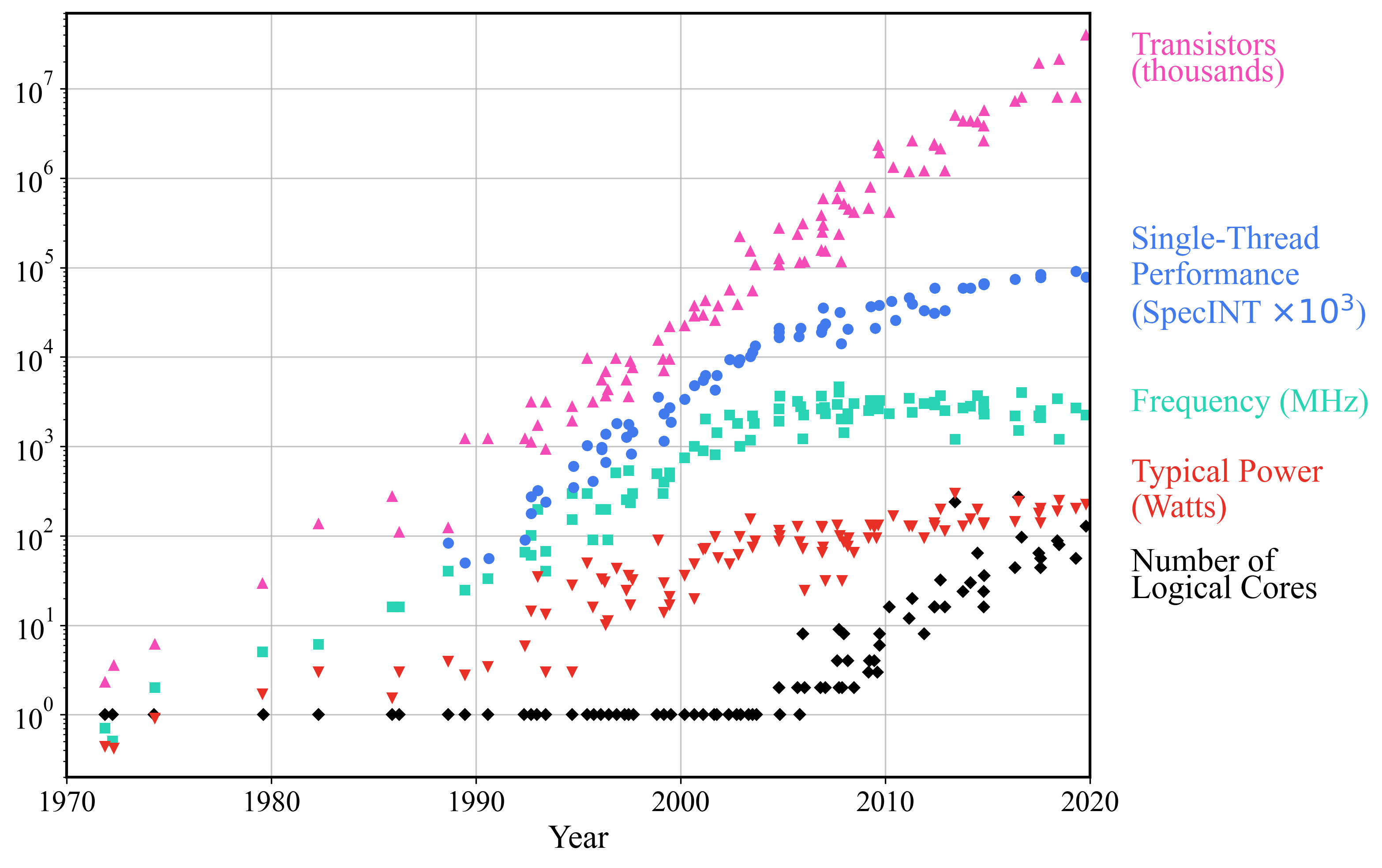}\\
  \footnotesize\textbf{(a)}
\end{minipage}
\hfill
\begin{minipage}{0.45\textwidth}
  \centering
  \includegraphics[width=\linewidth]{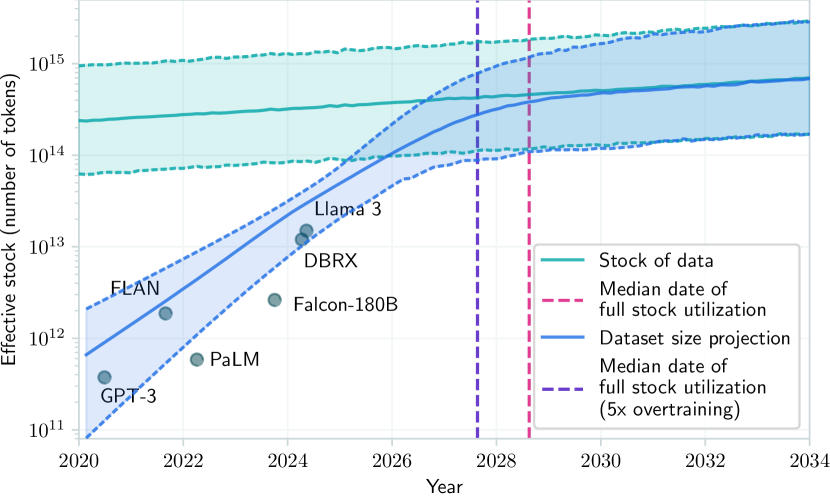}\\
  \footnotesize\textbf{(b)}
\end{minipage}

\vspace{1ex}

\begin{minipage}{0.45\textwidth}
  \centering
  \includegraphics[width=\linewidth]{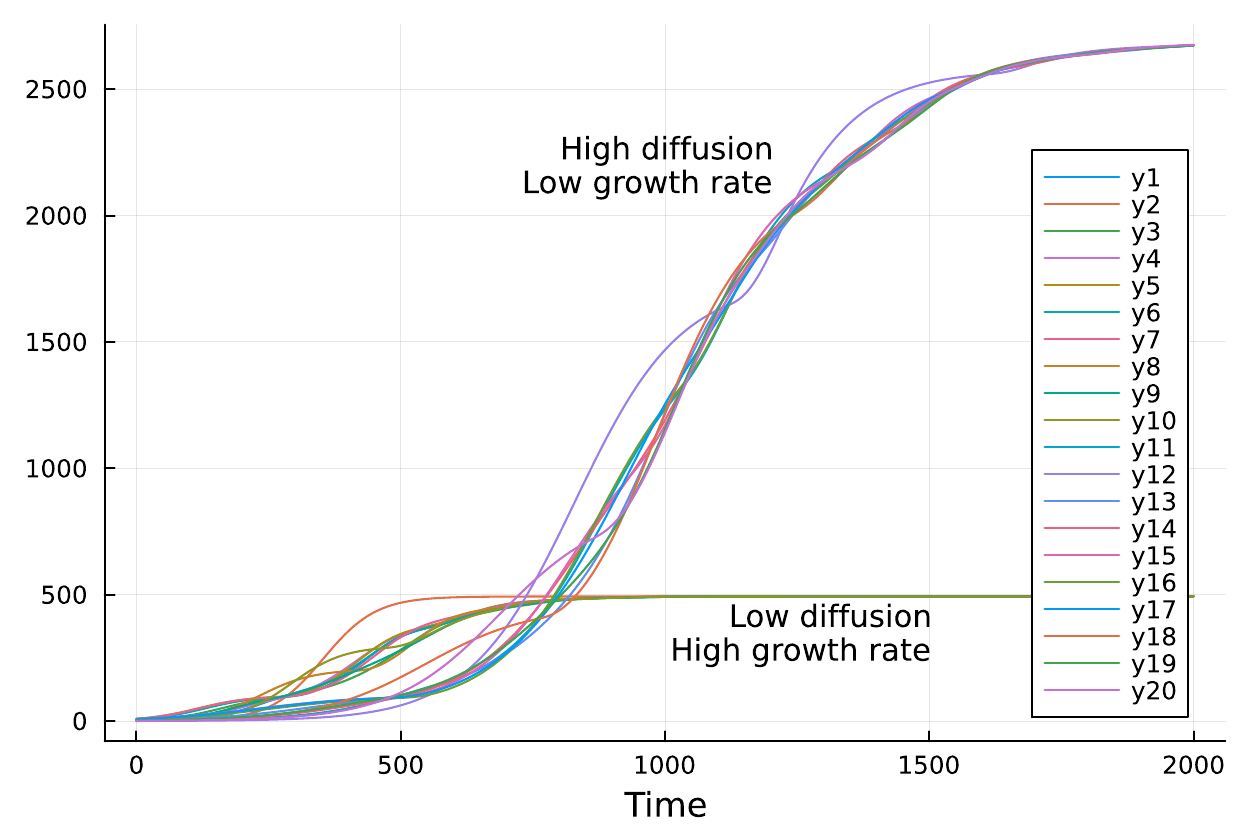}\\
  \footnotesize\textbf{(c)}
\end{minipage}
\hfill
\begin{minipage}{0.45\textwidth}
  \centering
  \includegraphics[width=\linewidth]{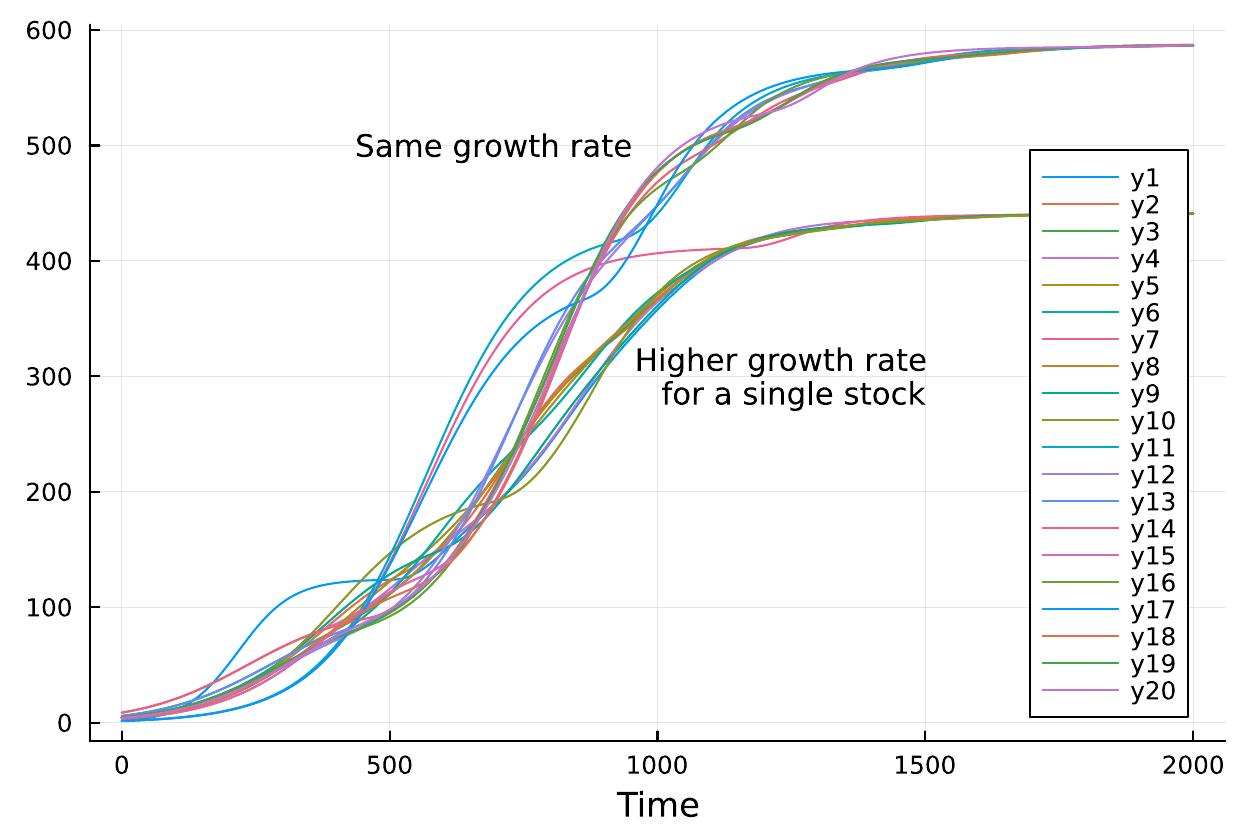}\\
  \footnotesize\textbf{(d)}
\end{minipage}

\caption{Evolution of (a) hardware performance in computing technologies \citep{rupp2021microprocessor}. The figure is generated independently from the original data collected in 2010 by M. Horowitz, F. Labonte, O. Shacham, K. Olukotun, L. Hammond, and C. Batten, showing long-term trends in transistor counts, performance, frequency, power, and cores, and (b) software-driven progress in LLMs \citep{villalobos2024willwerunoutofdata}, illustrating the rapid acceleration of AI capabilities. Technological growth through innovation diffusion in conditions of (c) high growth rates and low diffusion and inversely, low growth rates and high diffusion, and (d) a single stock with higher growth rate.}
\label{fig:1}
\end{figure}

The parameters, listed in Table \ref{table:1}, were chosen to illustrate the typical behaviors exhibited by the models and are not intended to be a faithful calibration to real systems. However, the structure of the models and their qualitative behavior (steady states, oscillations, chaos) in specific circumstances are meant to capture dynamics similar to that seen in the real world, such as for technological development and NFTs trends in transaction activity, see Figs. \ref{fig:1}.

\setlength\extrarowheight{4pt}
\begin{table}[t]
\centering
\begin{tabular}{|l|l|l|}
\hline
Parameter & Meaning & Value \\
\hline
$b$ & Investment into supply rate & $2.5\times 10^{-2}$ \\
\hline
$d$ & Production decay rate & $5\times 10^{-2}$ \\
\hline
$r$ & Demand growth/recovery rate & $10^{-2}$ \\
\hline
$\alpha_{1}$ & Demand satisfaction rate & $5\times 10^{-3}$ \\
\hline
$\alpha_{2}$ & Investment allocation rate & $5\times 10^{-3}$ \\
\hline
$c$  & Investment decay rate & $5\times 10^{-2}$ \\
\hline
$\sigma$ & Diffusion rate & $10^{-2}$ \\
\hline
$x_{0}$ & Initial supply & $1$ \\
\hline
$y_{0}$ & Initial demand & $0-10$ \\
\hline
$z_{0}$ & Initial investment & $1$ \\
\hline
$u_{0}$ & Initial market size & $100$ \\
\hline
\end{tabular}
\caption{\label{tab:example} Parameter definitions and baseline values used in the diffusion and market dynamics models.}
\label{table:1}
\end{table}

\section{Results}

\subsection*{Pure diffusion model}

Firstly, as a common feature across different scenarios we can see the general dynamics of the model, according to which technologies grow but reach a saturation point. However, due to higher development of other technologies the stagnated technologies can restart growing. In Fig. \ref{fig:1}(c) we contrast two cases: the first where the growth rate is set to a high value and the diffusion rate is set to a low value, and the second is the opposite situation, where the diffusion rate is higher and the growth rate is lower. We notice that low growth rates and high diffusion rates lead to sustained growth for a much longer time period compared to the high growth/low diffusion case. In addition, a significantly higher level of development is reached for all the technologies. The further illustrate this point, in Fig. \ref{fig:1}(d) we only increased the growth rate of a single technology and still notice a noticeable decline in overall development.

Hence, somewhat surprisingly, high growth rates and low diffusion lead to faster stagnation at a lower development level for the interconnected technologies. Implications of these dynamics are expanded upon in the Discussion section.

\begin{figure}[t]
\centering

\begin{minipage}{0.45\textwidth}
  \centering
  \includegraphics[width=\linewidth]{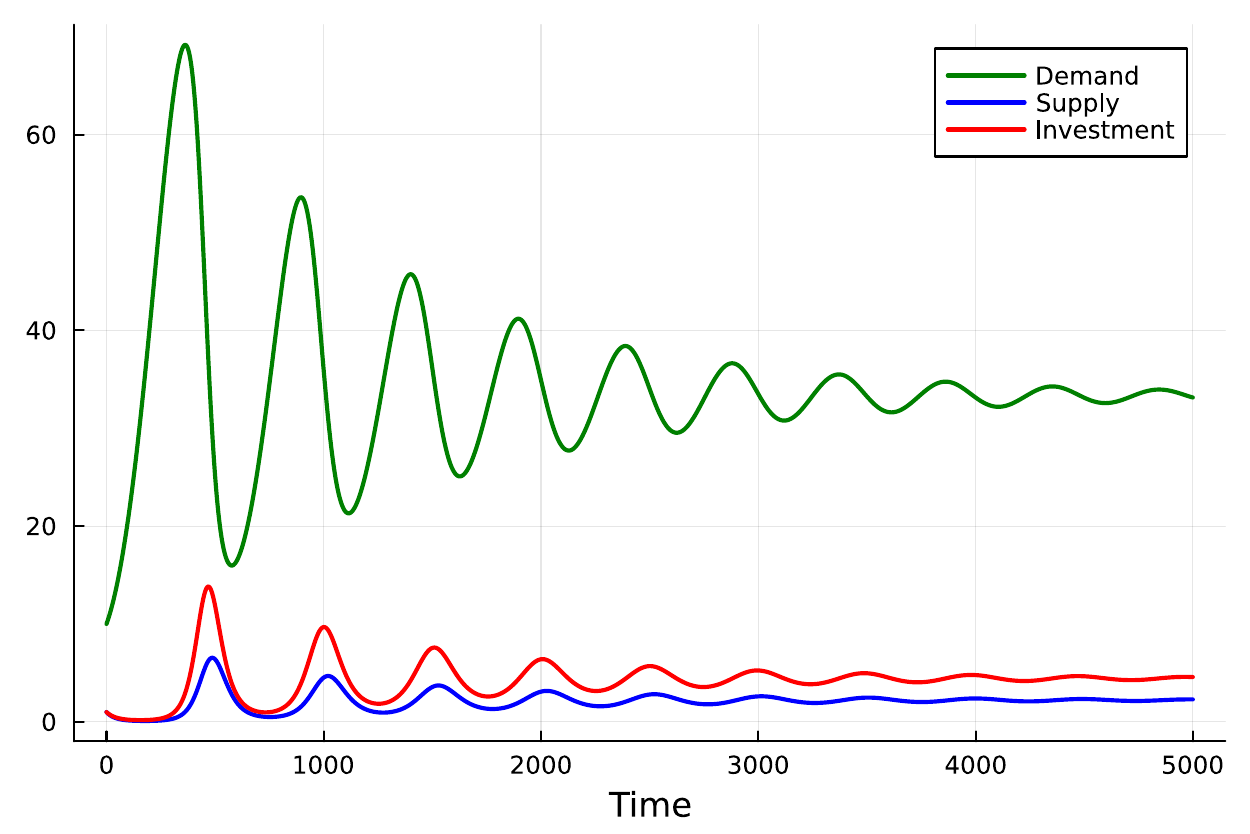}\\
  \footnotesize\textbf{(a)}
\end{minipage}
\hfill
\begin{minipage}{0.45\textwidth}
  \centering
  \includegraphics[width=\linewidth]{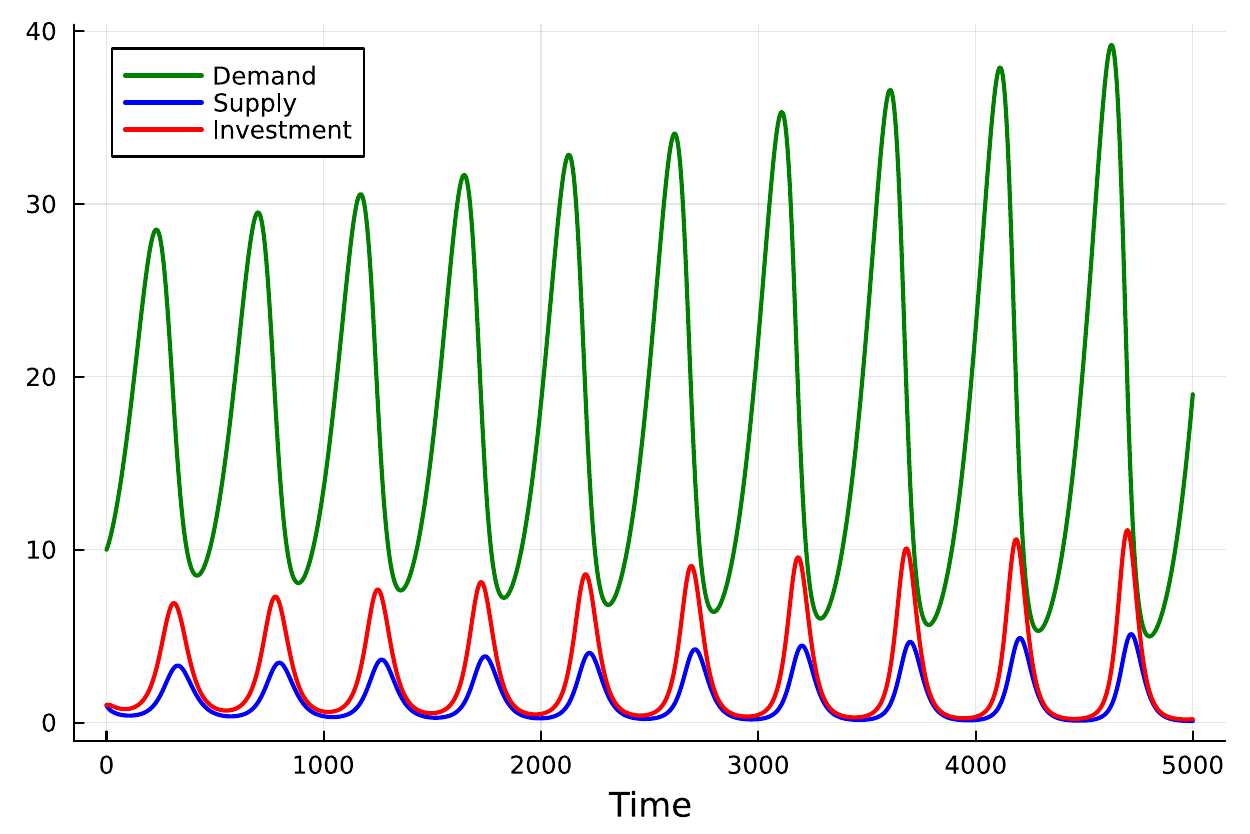}\\
  \footnotesize\textbf{(b)}
\end{minipage}

\vspace{1ex}

\begin{minipage}{0.45\textwidth}
  \centering
  \includegraphics[width=\linewidth]{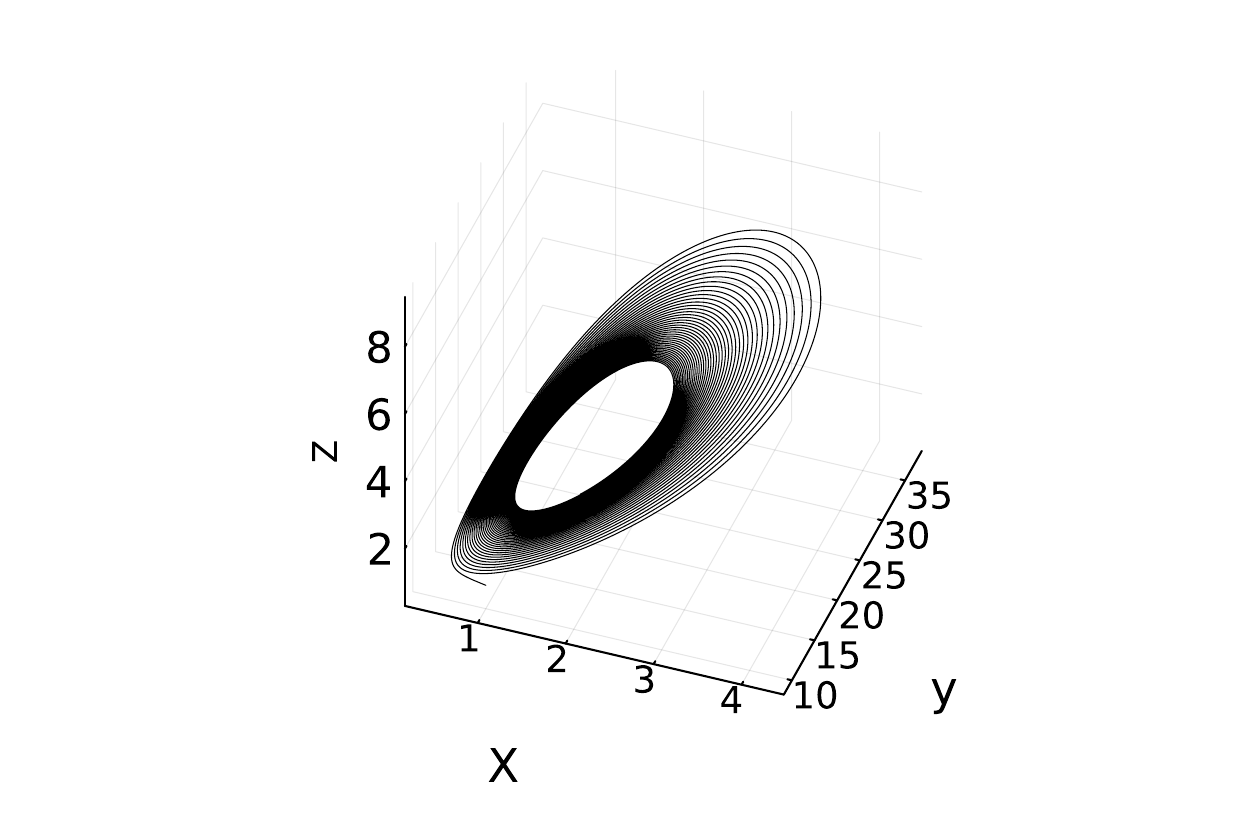}\\
  \footnotesize\textbf{(c)}
\end{minipage}
\hfill
\begin{minipage}{0.45\textwidth}
  \centering
  \includegraphics[width=\linewidth]{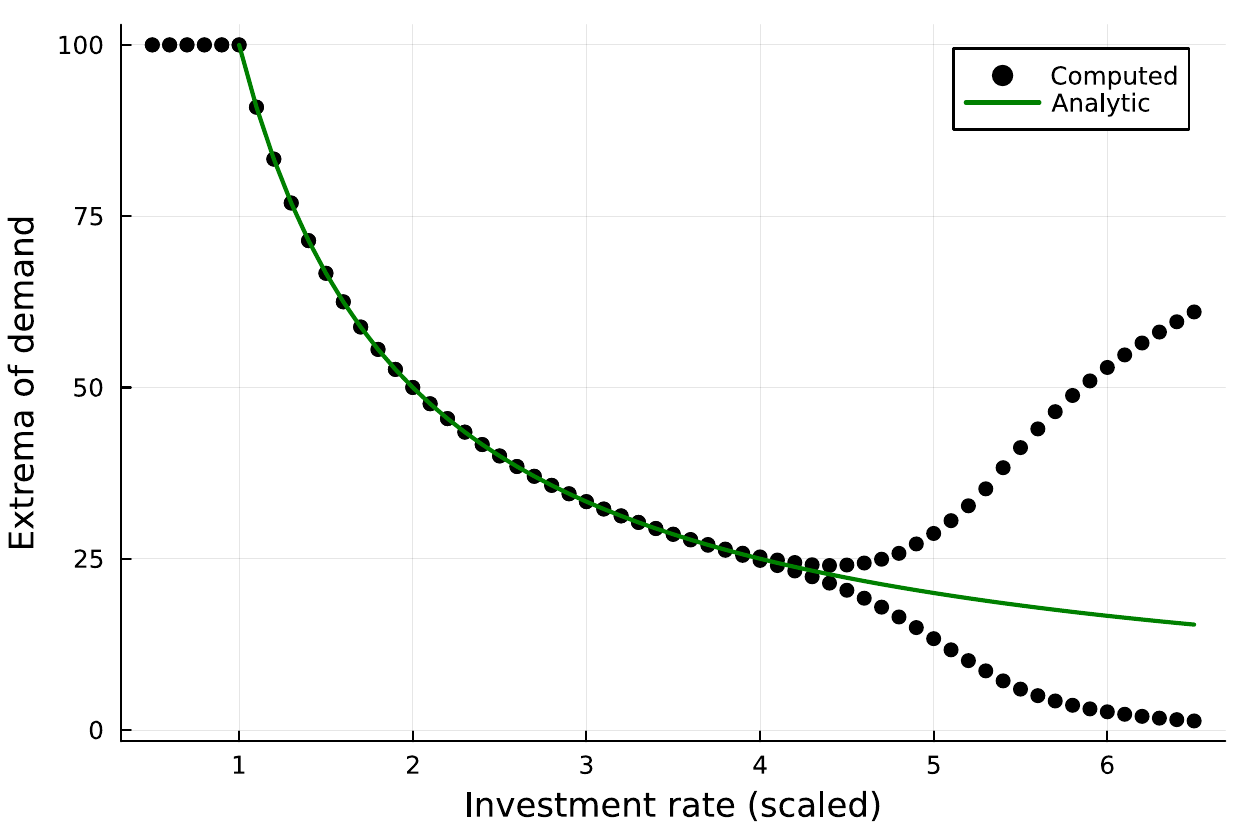}\\
  \footnotesize\textbf{(d)}
\end{minipage}

\caption{The dynamical behavior exhibited by the business cycle model \eqref{eq:2}: (a) reaches a fixed point (steady state) when $\alpha < \alpha_{c}$, (b) sustained oscillation when $\alpha > \alpha_{c}$. (c) Stable limit cycle and trajectory in phase space. (d) Bifurcation diagram (black dots) and equilibrium values for the demand (solid green).}
\label{fig:2}
\end{figure}

\subsection*{Business cycle model}

Fig. \ref{fig:2} illustrates the dynamical behavior of the business cycle model introduced in \eqref{eq:2}. When the investment parameter $\alpha$ is below a critical threshold $\alpha < \alpha_{c}$, the system stabilizes at a steady state, as shown in Fig. \ref{fig:2}(a). However, increasing $\alpha$ beyond this threshold induces persistent oscillations, as shown in panel Fig. \ref{fig:2}(b). Fig. \ref{fig:2}(c) shows these oscillations as a stable limit cycle in phase space, while Fig. \ref{fig:2}(d) presents the bifurcation structure: the system transitions from a stable fixed to a limit cycle as $\alpha$ increases. Together, these dynamics demonstrate how excess investment can destabilize markets, even in the absence of external shocks.

Depending on the parameter $\alpha$, there are in total three dynamical regimes of the system. Let $\alpha_{0} = \cfrac{cd}{b u_{0}} = 10^{-3}$ and:
\begin{equation}
\alpha_{c} = \frac{c^{2}+3cd+d^{2}+\sqrt{c^{4}+6c^{3}d+11c^{2}d^{2}+4c^{2}dr+6cd^{3}+4cd^{2}r+d^{4}}}{2bu_{0}} = 5\times 10^{-3}
\label{eq:ac}
\end{equation}
If $\alpha < \alpha_{0}$ then the investment rate is too low and the demand is not being satisfied, hence $x = 0$ and $y$ is maximum, see Fig. \ref{fig:2}(d). If $\alpha_{0} < \alpha < \alpha_{c}$ then the system reaches an interior stable equilibrium (fixed point), see Fig. \ref{fig:2}(a). If $\alpha > \alpha_{c}$ then sustained oscillations appear as a limit cycle, which can be interpreted as business cycles \citep{BENHABIB1979421}, see Fig. \ref{fig:2}(b). Hence, when increasing the $\alpha$ parameter the system undergoes a supercritical Hopf bifurcation leading to stable limit cycle, seen in Fig. \ref{fig:2}(c). The critical value is given by $\alpha_{c}$, which we determined analytically in equation \eqref{eq:ac}.

\begin{figure}[t]
\centering

\begin{minipage}{0.45\textwidth}
  \centering
  \includegraphics[width=\linewidth]{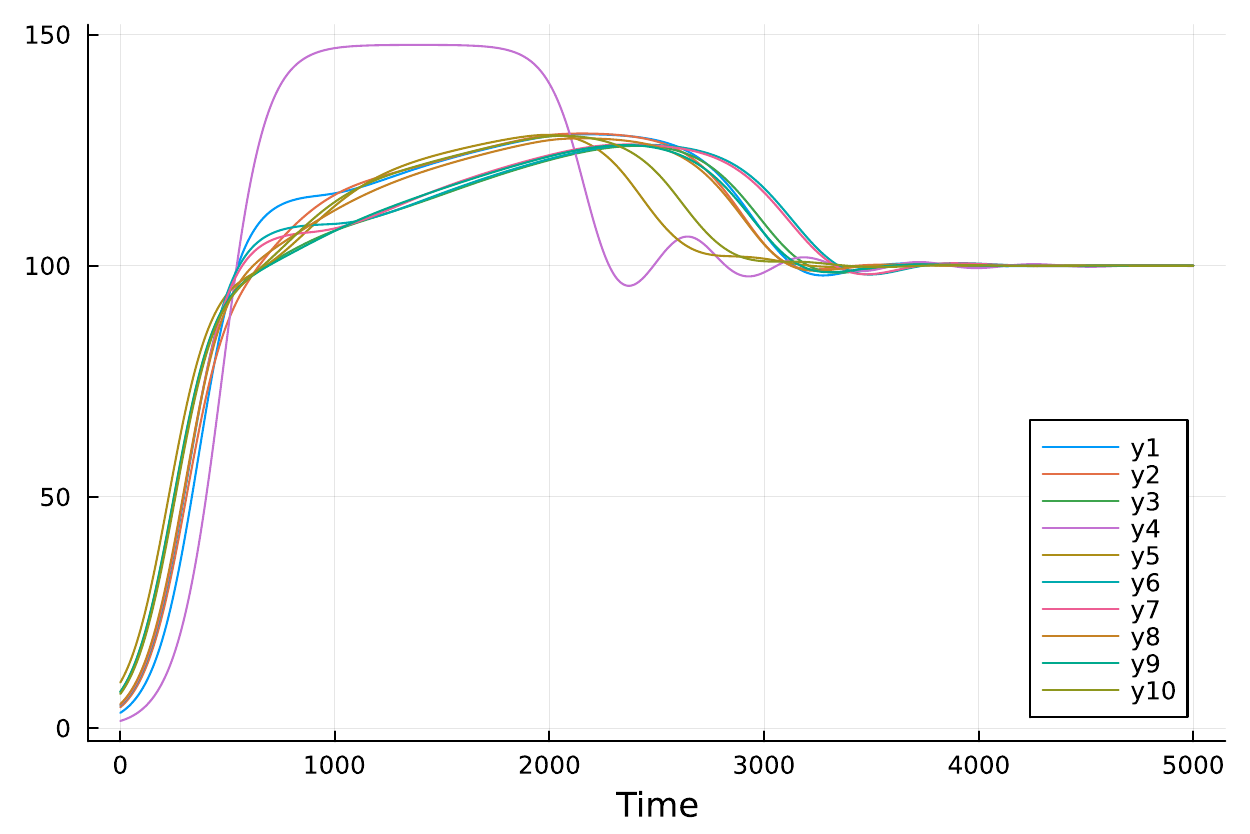}\\
  \footnotesize\textbf{(a)}
\end{minipage}
\hfill
\begin{minipage}{0.45\textwidth}
  \centering
  \includegraphics[width=\linewidth]{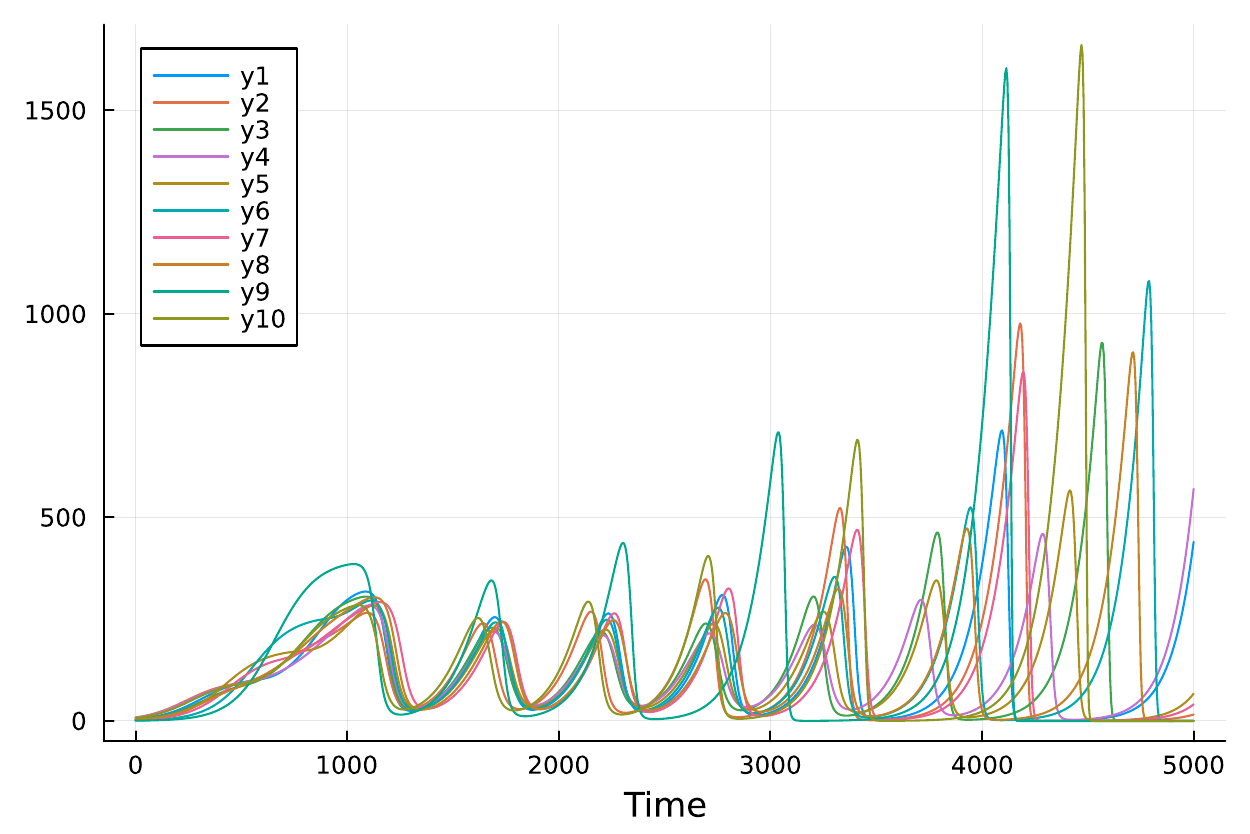}\\
  \footnotesize\textbf{(b)}
\end{minipage}

\vspace{1ex}

\begin{minipage}{0.45\textwidth}
  \centering
  \includegraphics[width=\linewidth]{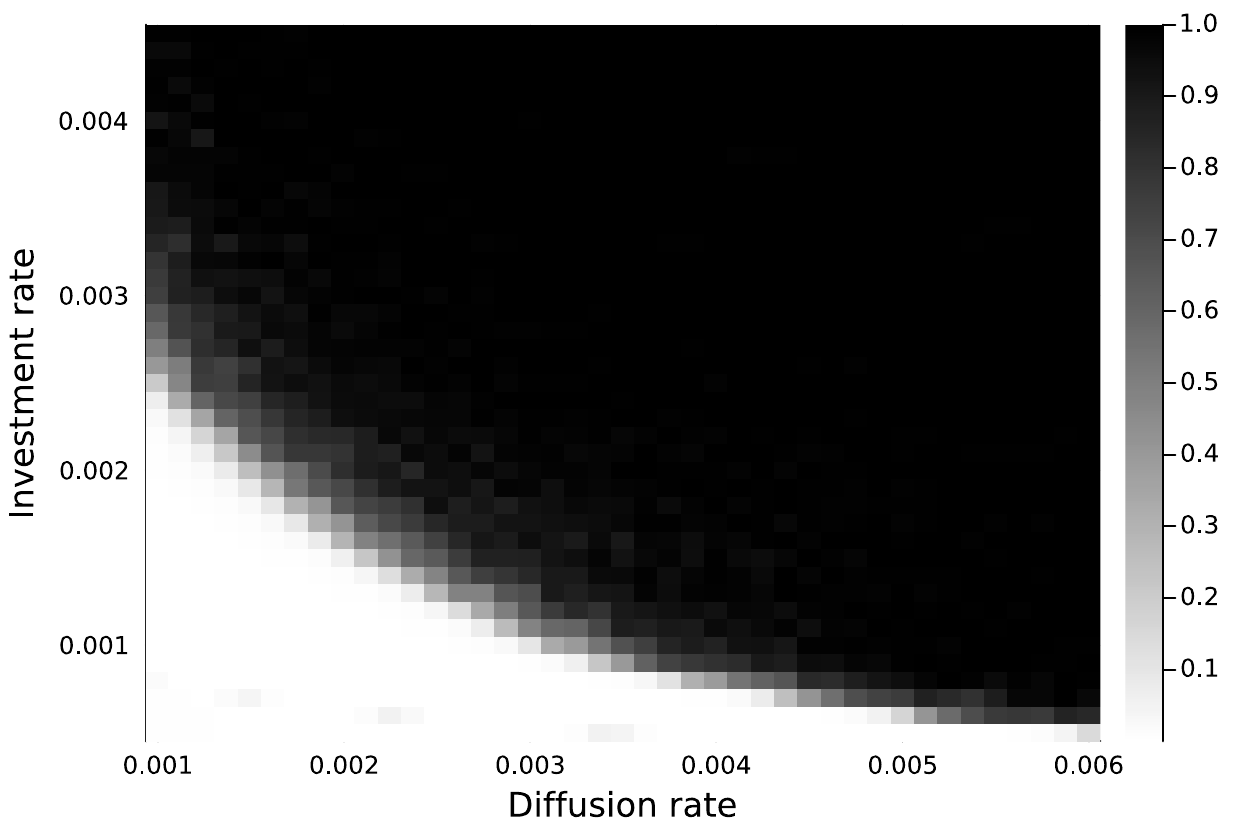}\\
  \footnotesize\textbf{(c)}
\end{minipage}
\hfill
\begin{minipage}{0.45\textwidth}
  \centering
  \includegraphics[width=\linewidth]{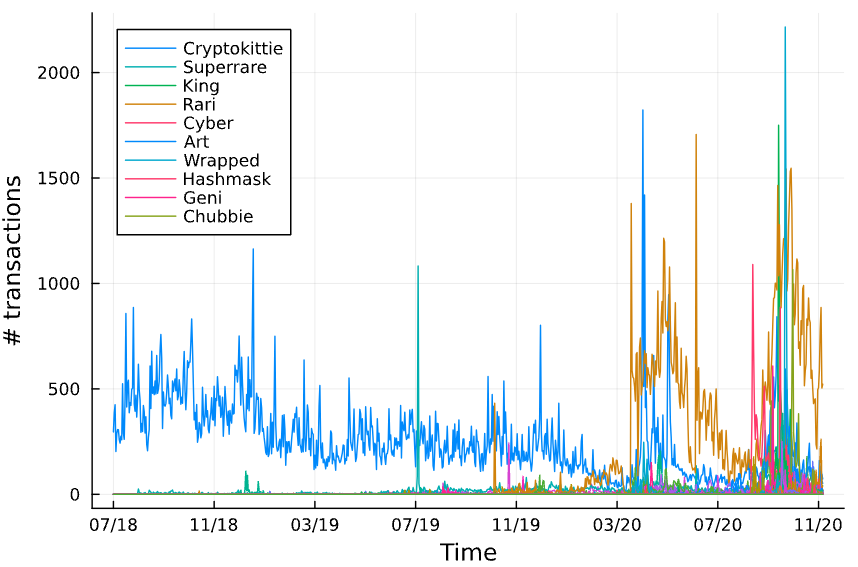}\\
  \footnotesize\textbf{(d)}
\end{minipage}

\caption{
Market model \eqref{eq:3} dynamics:  
(a) convergence to a steady state for $\alpha = \alpha_{0} = 10^{-3}$ and $\sigma = 10^{-3}$,  
(b) chaotic boom-bust cycles for $\alpha = 10^{-3}$ and $\sigma = 5\times 10^{-3}$,  
(c) steady-state regime (white) and chaotic boom-bust cycle (black). Qualitatively identical regime maps are obtained for quasi-regular and hub-dominated (scale-free-like) topologies; only the boundary shifts (lower in hub-dominated networks). (d) NFTs transactions data where the number of transactions serves as a proxy for demand.
}
\label{fig:3}
\end{figure}

\subsection*{Techno-landscape market model}

The results of the simulation of the combined diffusion and business cycles models are available in \ref{fig:3}. Specifically, it is possible to observed that, depending on the $\alpha$ and $\sigma$ parameters, we obtain two dynamical regimes: one where we get a stable fixed point, see Fig. \ref{fig:3}(a), and one where we obtain oscillatory behavior, in Fig. \ref{fig:3}(b). The single node dynamic dynamic in \eqref{eq:2} shows either a stable fixed point or a limit cycle with a clear periodicity, see Fig. \ref{fig:2}. In the network case modeled by \eqref{eq:3}, the dynamics for higher values of either $\alpha$ or $\sigma$ is chaotic with boom-bust cycles. This can be seen in Fig. \ref{fig:3}(c) where we plot the ratio $\cfrac{\max(\overline{y})-\min(\overline{y})}{\max(\overline{y})}$ where $\overline{y}$ is the average value of demands across the market. If the ratio is $0$ then the maximum and minimum value coincide, which occurs in a steady state. If the ratio is $1$, that indicates large oscillations where the maximum is much larger then the minimum. In Fig. \ref{fig:3}(d) we have included NFTs data whose patterns over time resemble those in Fig. \ref{fig:3}(b). We detail this comparison in the Discussion section.

\section{Discussion}

We have introduced and explore the behavior of three models that illustrate different behavior as relating to technological and business dynamics. On a more general perspective, all the models we have presented are theoretical in nature they nevertheless do reflect qualitative features seen in real data \citep{epstein2008model}. The model is built upon functional relations that assume causal links between the real-world properties represented by the variables. Our aim was to reproduce observed phenomena in a minimal way, providing a straightforward cause–effect explanation. The simplicity of this explanation directly reflects the deliberately elementary structure of the models \citep{edmonds2017different}.

In particular, the dynamics of the pure diffusion model reproduce the trends in technology, see Fig. \ref{fig:1}(a) and (b). The pure diffusion model can be thus considered representative of long-term, large-scale growth of the entire computer industry (and similar to other mature sectors) \citep{mukhopadhyay2009modelling}. 
On the other hand, the second model introduces instead business cycles, a well-known pattern in economic data \citep{kondratieff1979long}, and it aimed at generating such features in a simple manner. More specifically, we observed a chaotic dynamics in the full market model, with a quite erratic behaviour that match well the boom-bust patterns often seen in NFTs data \citep{nadini2021mapping}. We can interpret this to mean that investment or diffusion rates are high within the NFTs market, consistent with the characteristic hype of it \citep{baklanova2024investor}. Beyond this qualitative match, the model suggests that the observed NFT market dynamics can be understood as a case of transient chaos, where repeated overshoots ultimately lead to collapse. Also, it highlights the high sensitivity of the system to small variations in investment or diffusion rates, a feature consistent with the structural fragility of NFT markets; finally, the comparison shows that while such digital markets unfold on much shorter time scales, they share the same underlying mechanisms observed in longer-term technological sectors.
While there has been other boom-bust events, notably the ".com" and AI winter of the 80s \citep{menzies200321st}, the NFTs time scale is much shorter, which facilitates data collection and more direct comparison. While the results shows that diffusion and investment may induce transient boom–bust dynamics even in the absence of paradigm crises, we do not suggest inevitability; rather, we delineate conditions in which the likelihood of winters increases.

\begin{figure}[t]
\centering

\begin{minipage}{0.45\textwidth}
  \centering
  \includegraphics[width=\linewidth]{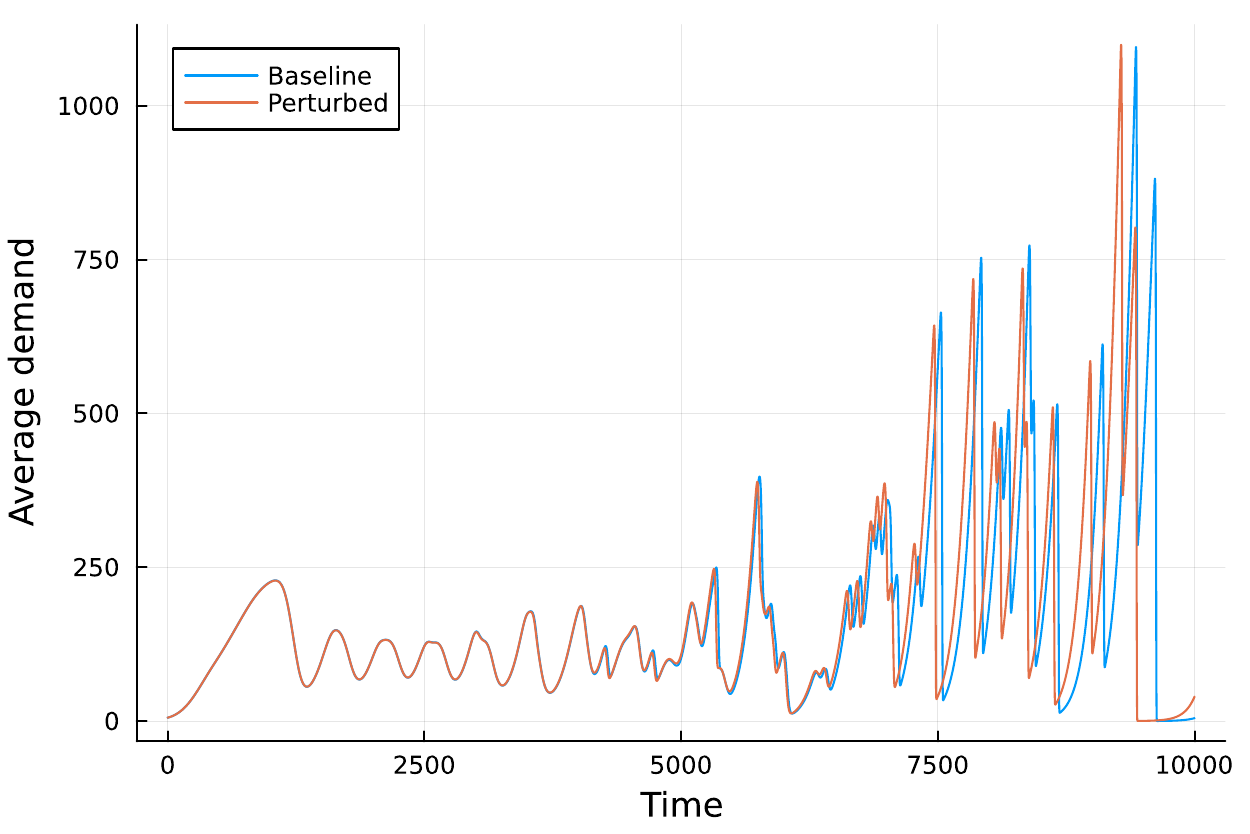}\\
  \footnotesize\textbf{(a)}
\end{minipage}
\hfill
\begin{minipage}{0.45\textwidth}
  \centering
  \includegraphics[width=\linewidth]{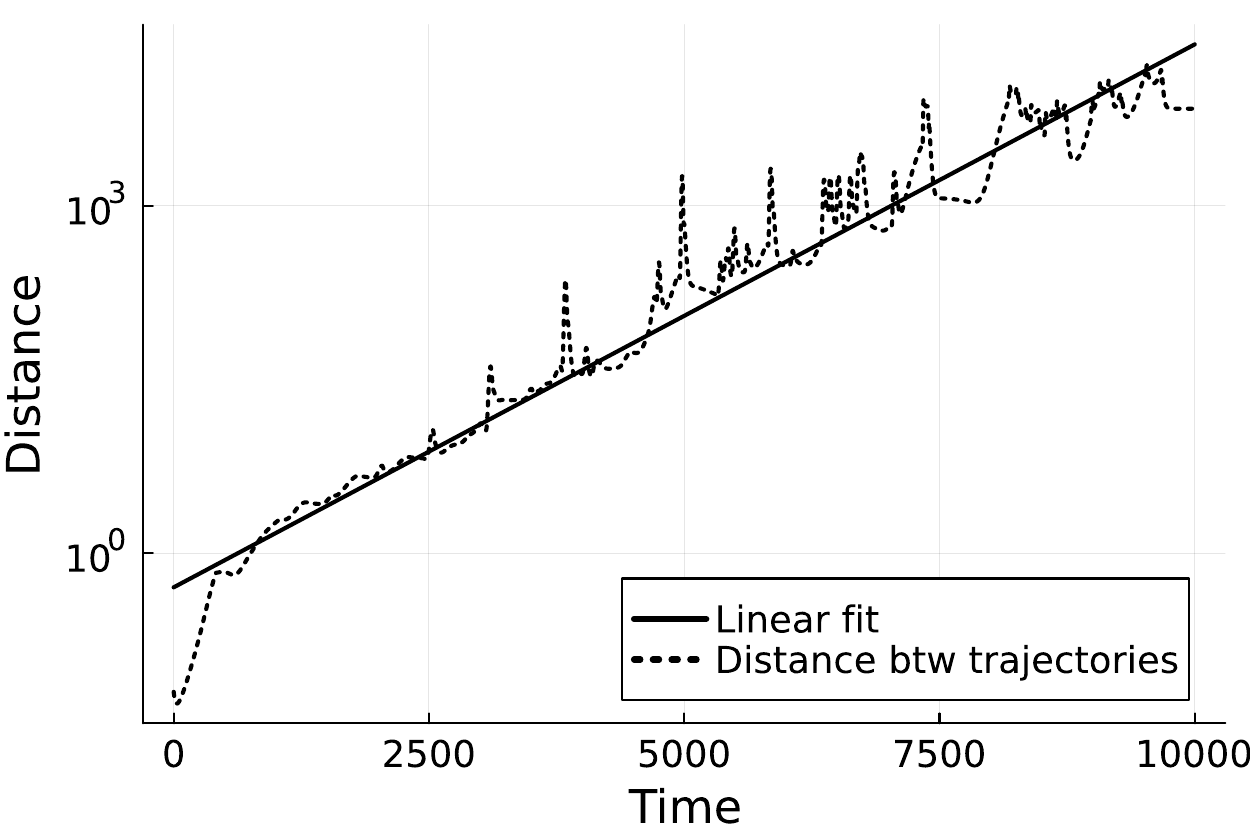}\\
  \footnotesize\textbf{(b)}
\end{minipage}

\caption{
(a) Plot of demand averages with different initial conditions, which differ by $10^{-2}$. (b) Computation of Lyapunov exponent with a positive value of $10^{-3}$, showing an exponentially growing distance between the trajectories. This indicates a chaotic regime for $\alpha = 10^{-3}$ and $\sigma = 5\times 10^{-3}$.
}
\label{fig:4}
\end{figure}

In general, a qualitative comparison can be drawn between the simulated dynamics and the empirical datasets considered. In the case of microprocessors, the plateau in single-thread performance and clock frequency, followed by renewed growth through multicore architectures, mirrors the role of diffusion ($u_i$) in lifting stagnated trajectories ($y_i$). For LLMs, phases of rapid progress could be interrupted by compute or data bottlenecks, after which advances in adjacent domains (e.g., data practices, deployment infrastructure) effectively increase $u_i$ and enable the next phase of expansion. What we observe now in the data is, for obvious reason, the first part of this trend. Finally, NFTs display highly asymmetric boom–bust cycles, with sharp collapses and increasing amplitude in early stages, closely resembling the transient-chaos regime of the model under high diffusion ($\sigma$) or investment ($\alpha$).

The following discussion now turn to an analysis of the observed behavior of the three models. 

Firstly, we considered a model for the diffusion of technological innovation. The model demonstrates that, counter to common business incentives, higher intrinsic growth rates may lead to earlier stagnation unless diffusion mechanisms are in place, when exogenous effects from outside the market are not taken into consideration \citep{biswas2015role}. In particular, the high growth/low diffusion scenario can be considered illustrative of a market with strong endogenous incentives for corporate growth (with more takeovers and merges, hence fewer and bigger companies) and protective/restrictive legislation on intellectual properties rights (less sharing of knowledge, copyright and patent rights persisting for long periods of time). An illustrative example of this regime is the development of AI in the private market, which can be structurally predisposed to unsustainable trajectories and premature saturation. 
The low growth/high diffusion environment is one where the growth of enterprises is not a priority and knowledge sharing and diffusion of findings is favored, with open practices regarding code, data and findings \citep{du2023exploring}. The academic environment with universities and research institutes more closely aligns with this regime than companies \citep{cheah2020effect}.

Secondly, we introduced a business cycle model whose structure captures two aspects of economic dynamics. On one hand, the evolution of production over time and the associated costs are directly measurable and can be tracked \citep{sterman2015system}. This informs the structure of the supply and investment variables. However, the dynamics of demand, the way supply satisfies it and the impact of marketing and investment strategies are not directly observable or measurable and can vary drastically across different industries \citep{islam2022current, abdulkareem2021promise}. For example, in oil exploitation, short-term production increases are prioritized \citep{costa2010value} while in the pharmaceutical and automotive industry investments and returns are planned over the long-term \citep{montoya2010dynamic, laabs2010long}. In the case of AI, certain applications such as language models exhibit rapid adoption \citep{polyportis2024longitudinal, hilling2025imperative} and the actual market signals are often driven by hype cycles, investor sentiment, and speculative expectations rather than grounded, measurable consumer needs \citep{kieffer2023learning, LaGrandeur2023The, Ahmadirad2024Evaluating}. Furthermore, the speed of product iteration and the lack of standardized performance metrics across use cases make it difficult to assess true demand saturation or long-term value creation \citep{cavalcanti2024exploring, Kwa2025Measuring}, adding  a layer of endogenous unpredictability.
While the proposed business cycle model has certain debatable elements (depending on what industry is being considered), to the best of our knowledge it represents the simplest three dimensional quadratic system with realistic dynamics of supply, demand and investment. As such, it can form both a research and pedagogical tool as it allows for analytic treatment and furthermore, from an economics perspective, it moves beyond the focus on market equilibrium to a more dynamic picture \citep{buldu2019taming}.

Lastly, we have build a model of diffusion of technological innovation incorporating business dynamics. Provided that the investment and diffusion rate parameters have lower values, as in the bottom-left corner of Fig. \ref{fig:3}(c), then we arrive at a steady state. However, in contrast to the pure diffusion model, there is an initial overshoot followed by a decrease that eventually leads to the equilibrium \citep{fenn1995hypecycle}, see Fig. \ref{fig:3}(a). If the investment rate is increased, then oscillations appear as in the simple business model, but the critical threshold is now lower. Furthermore, keeping the investment rate constant and increasing the diffusion rate also leads to a boom-bust regime, where chaotic dynamics emerges over time, see Fig. \ref{fig:3}(b). However, the chaotic regime is transient and eventually the demand collapses to zero \citep{nadini2021mapping, roman2023collapse, roman2025maximum}.

To make the instability explicit, Fig.~\ref{fig:4}(a) reports two runs of the networked market model~(3) whose initial conditions differ by a small perturbation of the demand state. The observable shown is the network average demand, $\bar{y}(t)=\frac{1}{N}\sum_i y_i(t)$. Despite near-identical starts, the trajectories separate rapidly when parameters are in the boom--bust region of Fig.~\ref{fig:3}(c) (e.g., $\alpha=10^{-3}, \sigma=5\times10^{-3}$; cf.\ Fig.~\ref{fig:3}b), indicating sensitivity to initial conditions. In Fig.~\ref{fig:4}(b) we quantify this by the standard two-trajectory estimate of the (finite-time) Lyapunov exponent $\lambda$, obtaining a positive value $\lambda \approx 10^{-3}$ over the time window shown. A positive exponent implies local exponential divergence of nearby trajectories, a hallmark of deterministic chaos, and is consistent with the irregular boom--bust oscillations seen in Fig.~\ref{fig:3}(b). Together with the regime map in Fig.~\ref{fig:3}(c), these diagnostics support our characterization of the high-$\sigma$/high-$\alpha$ region as transiently chaotic.

We repeated the analysis of Fig.~\ref{fig:3} while keeping the number of nodes $N$ and the mean degree fixed, comparing (i) quasi-regular graphs with a narrow, approximately Gaussian degree distribution (e.g., random-regular) to (ii) hub-dominated networks with heavy-tailed, power-law degree distributions (scale-free-like). The two-regime picture in Fig.~\ref{fig:3}(c) persists in both cases: a stable region at low diffusion/investment and a boom--bust region with irregular oscillations once coupling is strong enough. Degree heterogeneity shifts the threshold: hub-dominated networks lower the transition boundary in either $\sigma$ or $\alpha$, so chaotic transitions appear at weaker coupling. Intuitively, highly connected hubs synchronize and amplify local booms, making the system more sensitive.

From a qualitative perspective, we can conceptualize policy/managerial interventions as changes to the effective network topology, both the baseline coupling $\sigma$ and the degree heterogeneity. Open standards, interoperable interfaces, and data portability increase connectivity while enabling many-to-many links; when they disperse centrality (flattening the degree distribution), they raise the threshold for boom--bust and moderate cycle amplitude by limiting hub-driven amplification. By contrast, platform consolidation or winner-take-most procurement concentrates links on a few hubs, lowering the boundary in Fig.~\ref{fig:3}(c) and making chaotic transitions more likely. Consortia and shared infrastructure can act as ``bridges'' that shorten paths and improve coordination; designed to avoid single points of control, they damp oscillations via risk-sharing and capacity smoothing, whereas highly centralized consortia risk reproducing hub-dominance effects.

The two regimes, of either approaching a steady state or chaotic, can also be compared in terms of the market capacity $u$, looking at how it evolves over time. In the stable regime the market capacity reaches a steady state as well, whereas in the chaotic case the capacity keeps increasing. While the market capacity grows larger, this does not imply boundless growth, rather it means that the demand recovers quickly which spurs a new a rapid growth in investment and supply; the race to satisfy the demand leads to a boom and ultimately a bust. While the underlying market capacity remains large it harbors an inherent volatility in any attempt to tap into it: every subsequent boom is generally larger than prior ones and similarly for the bust. This reflects an old adage \citep{tainter1988collapse, wright2004short}: "Each time history repeats itself, the price goes up."

These dynamics are particularly relevant when considering the future of AI \citep{oosthuizen2022fourth}. The model shows that a persistently growing market capacity does not equate to sustainable development, and that it can create the illusion of limitless opportunity, encouraging over-investment and speculative expansion \citep{bloom2020ideas}. As each cycle amplifies both the boom and the bust, the system becomes increasingly more fragile at each iteration. In such a context, a sudden collapse in demand, driven by disillusionment, regulatory shocks, or technological stagnation, could trigger a new AI winter \citep{hendler2008avoiding}. From a policy perspective, this calls for mechanisms that moderate the pace of investment and enhance diffusion across technological domains, which, due to short-term economic and geo-political competition, is exactly the opposite of what is happening right now. Reducing the investments rates, support open standards, interoperability, and knowledge-sharing practices may help distribute innovation. Moreover, counter-cyclical public funding—supporting foundational research during hype downturns—could stabilize long-term trajectories. Without such interventions, the AI sector risks repeating the historical pattern of overshoot and collapse, but with broader consequences for economic stability, public trust, and innovation ecosystems.

Moreover, it is relevant to discuss the relationship between our results and the AI Winter of the 1980s. While we acknowledge that historical winters were not caused exclusively by economic dynamics, but also by deeper cognitive and paradigmatic crises, our model captures structural mechanisms that contributed to the downturn and thus provides a complementary perspective. First, our results show that excessive investment amplifies both booms and busts. This hype–investment cycle echoes the surge of expectations around expert systems \citep{gill1995early}, where heavy industrial and governmental funding created unsustainable growth that collapsed once limitations became clear. Second, we show that high intrinsic growth combined with low diffusion accelerates stagnation. In the 1980s, research was concentrated in a few firms and laboratories \citep{Grimson1987AI}, with limited openness and standardization \citep{phillips1999if}, reproducing the unsustainable trajectories reflected in our model. Third, we highlight how strong interdependencies across technologies can generate systemic fragility: the AI ecosystem in the 80s was overly dependent on a narrow symbolic paradigm \cite{nilsson2009quest}, which made it vulnerable when that paradigm faltered. Finally, our business cycle model emphasizes the volatility of demand, shaped by hype and reputation rather than stable needs. This mirrors the speculative character of the expert systems market \cite{goldstein1995uncertainty, harmon2022expert}, which collapsed once performance failed to meet expectations.

\section{Conclusion}

This study introduces a set of interconnected dynamical models to better understand the diffusion of technological innovation and the economic forces shaping its trajectory. By integrating models of innovation diffusion, business cycles, and networked market dynamics, we capture a range of behaviors from stable growth to oscillatory and chaotic regimes.

More specifically, the results of the three models can be summarized as follows. The pure diffusion model reproduces the long-term stagnation and renewal cycles observed in mature technologies, showing that higher intrinsic growth combined with low diffusion leads to earlier stagnation, while lower intrinsic growth combined with stronger diffusion mechanisms supports longer and more sustainable development trajectories. The business cycle model captures the transition from stability to oscillations as investment surpasses a critical threshold, reproducing the endogenous emergence of cycles well documented in economic theory. Finally, the combined techno-landscape model demonstrates that when both diffusion and investment rates are high, transient chaotic boom–bust cycles appear. Importantly, this behavior is robust to network topology, holding network size and mean degree constant, the same two regime structure, namely a stable regime at low diffusion and investment and an irregular boom-bust regime once coupling is sufficiently strong, emerges in both quasi-regular and hub-dominated networks. Higher degree heterogeneity lowers the diffusion and investment thresholds for chaos, as highly connected hubs synchronize and amplify local booms, thereby increasing overall system sensitivity. This finding is consistent with empirical patterns such as those observed in the NFT market and suggests that structurally similar dynamics may underlie technological fields such as AI.

Notably, our findings suggest that high investment or diffusion rates, typically seen as drivers of innovation, can lead instead to instability and boom–bust dynamics. These outcomes mirror real-world patterns in rapidly evolving sectors like NFTs and raise caution for areas such as AI development, where hype and aggressive funding often outpace sustainable growth. Our analysis indicates that diminishing returns, architectural limits, and data constraints may lead to a downturn in AI investments, echoing structural patterns of past cycles. While not predictive in a temporal sense, the model highlights generative mechanisms that make an approaching “AI winter” structurally plausible, even if its timing and severity remain uncertain. The connection with the historical AI winter of the 1980s is particularly relevant: although that downturn was also driven by deeper cognitive and paradigmatic crises, our models capture complementary structural dynamics, such as excessive investment, limited diffusion, and systemic fragility, that reinforce the likelihood of collapse. From a broader perspective, these insights carry implications for both research and policy. Open standards, interoperability, and knowledge-sharing practices can help mitigate premature stagnation by strengthening diffusion, while counter-cyclical public funding may stabilize long-term trajectories by supporting foundational research during downturns. Future empirical work could further test these mechanisms, for example by integrating maturity indicators such as TRL, MRL, or CRL as dynamic inputs or calibration points, and by exploring the interaction of structural cycles with cognitive and semantic aspects of AI development.

In conclusion, the models presented here provide a conceptual foundation for understanding how technological diffusion and investment interact to shape long-term trajectories. They highlight that, under present conditions, a new AI winter is not inevitable but remains structurally plausible. Recognizing and moderating the underlying mechanisms may therefore be essential to fostering more resilient technological ecosystems.


\section*{Conflict of Interest Statement}

The authors declare that the research was conducted in the absence of any commercial or financial relationships that could be construed as a potential conflict of interest.

\section*{Author Contributions}

S.R. contributed with conceptualization, investigation, methodology, analysis of results, project administration, software, writing – original draft, writing – review and editing, F.B. contributed with data curation, resources, methodology, software, validation, writing – review and editing.

\section*{Funding}
We would like to thank Open Philanthropy for supporting this work.

This publication is supported by the European Union's Horizon Europe research and innovation programme under the Marie Sk\l{}odowska-Curie Postdoctoral Fellowship Programme, SMASH co-funded under the grant agreement No. 101081355. The operation (SMASH project) is co-funded by the Republic of Slovenia and the European Union from the European Regional Development Fund.

\section*{Data Availability Statement}
The code to reproduce the results is available at: \url{https://doi.org/10.5281/zenodo.17335983}


\bibliographystyle{Frontiers-Harvard} 
\bibliography{biblio}




\end{document}